\documentclass{article}

\usepackage[colorlinks,urlcolor=blue,linkcolor=blue,citecolor=blue]{hyperref}
\usepackage[subrefformat=parens,labelformat=parens]{subcaption} 
\usepackage{color,array}
\usepackage{graphicx}
\usepackage{cite}
\usepackage{amsmath,amssymb,amsfonts}
\usepackage{tabularx} 
\usepackage{booktabs}
\usepackage{multirow}
\usepackage{multicol}
\usepackage[table]{xcolor}
\usepackage[margin=1in]{geometry}
\begin{document}

\title{Learning Incentive Structures for Cooperative Resilience in Multi-Agent Systems under Social Dilemmas}

\author{
    \small Manuela Chacon-Chamorro \and
    \small Luis Felipe Giraldo \and
    \small Nicanor Quijano 
    \thanks{This work was supported by Google through the Google Research Scholar Program, and by the UniAndes–DeepMind Scholarship 2023.}
    \thanks{The authors are with the School of Engineering, Universidad de los Andes, Bogotá, Colombia (emails: \{m.chaconc, lf.giraldo404, nquijano\}@uniandes.edu.co)}
    \thanks{Updated version submitted to IEEE Transactions on Computational Social Systems (TCSS). This preprint is under review for possible publication in IEEE.}
}

\date{\small}


\maketitle
\hrule
\begin{abstract}
Multi-agent social dilemmas, such as the tragedy of the commons, capture settings where individual incentives conflict with collective well-being, making these systems highly vulnerable to collapse under disruptions. In this context, this work studies \textit{cooperative resilience}, understood as the system-level ability to maintain collective well-being under perturbations through adaptive agent behavior. We propose a framework for learning incentive structures aligned with collective well-being in multi-agent reinforcement learning systems, where reward functions shape individual decision-making and collective behavior. A resilience metric is used to score and rank agent trajectories, allowing the inference of reward functions that promote resilient collective behavior. These inferred reward functions are integrated into the multi-agent reinforcement learning process to shape agent interactions in social dilemma settings. The approach is evaluated in resource-sharing environments subject to disruptions, using three incentive structures: individual incentives, resilience-aligned incentives, and a hybrid incentive structure that combines both individual and collective components. The results show that the hybrid incentive structure promotes sustained collective behavior, reduces collapse events associated with resource depletion, and preserves system performance under disruption. These findings highlight the role of incentive design as a mechanism for promoting resilient collective behavior and provide a computational framework for multi-agent social dilemmas under disruptions.

\end{abstract}

\vspace{0.2 cm}
\hrule

\section{Introduction}

\label{sec:introduction}

Multi-agent social dilemmas, such as the tragedy of the commons, capture settings where individual incentives conflict with collective well-being, often leading to inefficient or unstable system outcomes \cite{leibo2017multi, rios2023understanding, mu2024multi}. In these environments, agents must balance short-term individual gains with long-term collective objectives, giving rise to complex social dynamics and patterns of collective behavior. This tension becomes particularly critical in the presence of disruptions \cite{topolewicz2023impact, yildirim2019control}, where the system may not be able to maintain coordination and collapse due to over-exploitation or lack of adaptation.

A central challenge in such settings is the design of incentives that promote desirable collective behavior. In multi-agent reinforcement learning (MARL), agent behavior is shaped by reward functions, which implicitly define the incentives guiding decision-making. However, specifying reward functions that balance individual and collective objectives remains difficult, especially in mixed-motive environments where agents must coordinate under uncertain and dynamic social interactions \cite{hammond2025multi}. Poorly designed rewards can lead to selfish behavior, inefficient resource usage, or failure to sustain cooperation over time \cite{hughes2018inequity}.

These challenges motivate the study of \textbf{cooperative resilience}, a system-level property that characterizes the ability of a group of agents to maintain collective well-being under perturbations \cite{chacon2024cooperative,shraga2025collaboration}. Cooperative resilience captures how collective behavior and social dynamics evolve in response to disruptions, including the system's ability to absorb degradation and recover functionality. Despite its importance, the relationship between incentive design and cooperative resilience is still not adequately understood.

In this work, we address this gap by proposing a framework to \textit{learn incentive structures from behavior}. Rather than manually designing reward functions, we adopt an inverse reinforcement learning (IRL) perspective \cite{brown2019extrapolating, goktas2025efficient, ashwood2022dynamic, wu2022inverse}, where rewards are inferred from observed trajectories. In particular, we leverage a cooperative resilience metric to evaluate agent behavior at the trajectory level and induce preference rankings over trajectories. These rankings are then used within a preference-based framework to learn reward functions that align agent behavior with resilient collective outcomes. The inferred reward functions are integrated into Proximal Policy Optimization (PPO)-based multi-agent training and evaluated against PPO and QMIX baselines using conventional reward structures under multiple disruption scenarios.

We evaluate the proposed framework in a social dilemma inspired by the ``Commons Harvest'' environment of Melting Pot \cite{perolat2017multi, agapiou2022melting}. The results show that the learned incentive structures improve cooperative resilience, support long-term system survivability, and reduce collapse events, while maintaining high levels of resource utilization. Additional experiments in larger environments suggest that the approach can be extended to more complex multi-agent settings. 

Our approach differs from existing methods that promote cooperation through predefined reward shaping, interaction protocols, or social coordination mechanisms. Instead of prescribing how agents should behave, we infer the underlying incentive structures that give rise to desirable system-level properties. This perspective enables the discovery of reward functions that support resilient collective behavior without requiring explicit encoding of such behaviors. This work contributes a computational framework for learning incentive structures aligned with collective well-being, providing a new perspective on how reward design can be used to shape resilient collective behavior and social dynamics in multi-agent systems.

The remainder of this paper is organized as follows. Section~\ref{sec:background} reviews the background and related work. Section~\ref{sec:problem} presents the formulation of the problem. Section~\ref{sec:resilience} introduces the modeling of cooperative resilience. Section~\ref{sec:learning} presents the proposed framework for learning incentive structures from behavior. Section~\ref{sec:experimental} describes the experimental setting. Section~\ref{sec:results} presents the results of emergent collective behavior under disruptions. Section~\ref{sec:discussion} discusses the findings, and Section~\ref{sec:conclusions} concludes the paper.

\section{Background and Related Work}
\label{sec:background}
\subsection{Social Dilemmas and Collective Behavior}

Multi-agent social dilemmas, such as the tragedy of the commons and public goods dilemmas, describe situations where individual incentives are misaligned with collective welfare, often leading to inefficient or even catastrophic outcomes. In these settings, agents are driven to maximize their own payoff, while the collective optimum requires cooperative behavior, creating a fundamental tension between self-interest and the common good \cite{strumke2022social, mu2024multi}. This conflict has been extensively studied across disciplines, where it is shown that individually rational decisions can result in collectively suboptimal outcomes, including resource depletion, free-riding, and system collapse \cite{fatima2024learning}. As a result, understanding how cooperative behavior emerges, evolves, and can be sustained in such environments has become a central problem in the study of collective behavior in multi-agent systems \cite{mu2024multi, fan2024integral}.

Cooperative AI studies the design of multi-agent systems that achieve results that benefit the group as a whole \cite{dafoe2020open, hammond2025multi}. In mixed-motive environments, such as social dilemmas, the design of mechanisms that foster cooperation remains particularly challenging \cite{hammond2025multi}. Standard reinforcement learning approaches typically optimize individual rewards, which can encourage self-interested behavior and lead to the degradation of shared resources in social dilemmas \cite{rios2023understanding, strumke2022social, du2022reimagining, leibo2017multi}. As a result, agents may fail to coordinate effectively or overexploit common resources, giving rise to inefficient and unstable collective outcomes. These challenges are further exacerbated in the presence of disruptions, such as resource scarcity, unsustainable behaviors, or abrupt environmental changes \cite{jasper2004strategic, orner2025explaining}, increasing the difficulty of maintaining cooperation over time.

\subsection{Cooperative Resilience in Multi-Agent Systems}

The presence of disruptions, such as resource scarcity, unsustainable behaviors, or abrupt environmental changes, further complicates collective behavior in multi-agent systems \cite{jasper2004strategic, orner2025explaining}. Under these conditions, systems are particularly vulnerable to performance degradation or collapse, as agents struggle to sustain coordinated behavior over time.

In this context, resilience becomes essential to maintain collective well-being. \textit{Cooperative resilience} is defined \cite{chacon2024cooperative} as a system-level property, closely related to group resilience \cite{shraga2025collaboration}, which captures the ability of agents to maintain collective performance under perturbations. Unlike traditional notions such as robustness or stability, cooperative resilience captures the dynamic response of the system, including its ability to absorb, adapt, and recover from disruptions.

Building on approaches from ecology, infrastructure, and economic networks \cite{ayyub2014metrics, cimellaro2016peoples, gerges2022metricsIndex}, the methodology in \cite{chacon2024cooperative} provides a quantitative framework to evaluate this property. Specifically, resilience is assessed by comparing system performance under disrupted and baseline conditions with multiple indicators of collective well-being, resulting in a trajectory-level resilience score. This score enables systematic comparison between behaviors and provides a basis for analyzing how different interaction patterns and incentive structures influence the system's ability to maintain collective outcomes under stress.

\subsection{Incentive Design and Multi-Agent Learning}

A common approach to modeling decision-making in multi-agent systems is reinforcement learning (RL), where agents learn policies based on reward signals derived from interaction with the environment. In multi-agent settings, these reward signals play a central role in shaping agent behavior, as they implicitly define the incentives that guide individual decisions and influence collective outcomes. To address the challenges of coordination and cooperation, MARL methods such as QMIX \cite{QMIX} and COMA \cite{COMA} focus on learning decentralized policies while handling issues such as credit assignment. Complementary approaches incorporate explicit incentive mechanisms, including shaping social rewards \cite{jaques2019social, hughes2018inequity}, which promotes cooperation by accounting for the impact of an agent's actions on others.

More recent work has emphasized \emph{incentive design} as a central component in multi-agent systems. Mechanisms such as peer reward \cite{lupu2020gifting,yang2020learning}, norm formation through sanctions \cite{vinitsky2023learning}, and mutual recognition protocols \cite{phan2024emergent} allow agents to influence the rewards of others, encouraging pro-social behavior through structured interactions. These approaches highlight how carefully designed incentives can shape collective behavior. However, these methods typically assume that an appropriate reward structure is available. In mixed-motive environments, specifying such incentives is challenging, as rewards must balance individual objectives with collective well-being, particularly under disruptions. Manually designing these reward functions may fail to capture the system-level properties that govern desirable collective outcomes.

IRL offers an alternative by inferring reward functions from observed behavior \cite{adams2022survey, metelli2023towards, arora2021survey}. Classical IRL approaches often rely on optimal demonstrations, which are rarely available in complex environments \cite{brown2019extrapolating, goktas2025efficient, poiani2024sub}. In multi-agent settings, additional challenges arise due to joint action spaces and equilibrium considerations \cite{natarajan2010multi, littman1994markov, ccelikok2024inverse}, limiting scalability and applicability.

Preference-based IRL addresses these limitations by learning from comparisons between trajectories instead of assuming optimal behavior \cite{brown2019extrapolating, willis2025will}. It uses relative feedback, which is often easier to obtain in complex environments. This approach has also been extended to incorporate multiple types of feedback, providing a more flexible and informative signal for reward learning \cite{larian2025learner}. This perspective is well-suited for social dilemmas, where behaviors can be naturally ranked based on their collective outcomes. Such rankings provide a practical basis for inferring incentive structures aligned with desirable system-level properties and sustained collective behavior.

\section{Problem Formulation}
\label{sec:problem}

We consider multi-agent environments where agents interact under mixed-motive conditions and share access to common resources. The environment is modeled using the standard \emph{joint-state, joint-action} formulation of a Markov game, defined by the tuple $(S, A, P, R, \gamma)$, where $S$ is the global environment state, $A = A_1 \times \cdots \times A_n$ is the joint action space, $P$ is the transition function over the joint state--action space, $R$ denotes the reward structure, and $\gamma$ is the discount factor. This formalization is consistent with the multi-agent decision-process models presented in \cite{littman1994markov,boutilier1996planning}. Each agent executes its own decentralized policy $\pi_i(a_i \mid s)$.

In this setting, standard reward functions $R$ often prioritize short-term individual gains, which can undermine collective welfare, especially under disruptive conditions. This motivates our central objective: to learn a reward function that promotes cooperative resilience. To this end, we propose a two-step methodology: i) ranking trajectories using a cooperative resilience metric (see Section~\ref{sec:resilience}), and ii) learning a reward function from preferences using one of two methods—margin-based optimization or probabilistic modeling (see Section~\ref{sec:learning}).

\begin{figure*}[t]
    \centering
    \includegraphics[width=\linewidth]{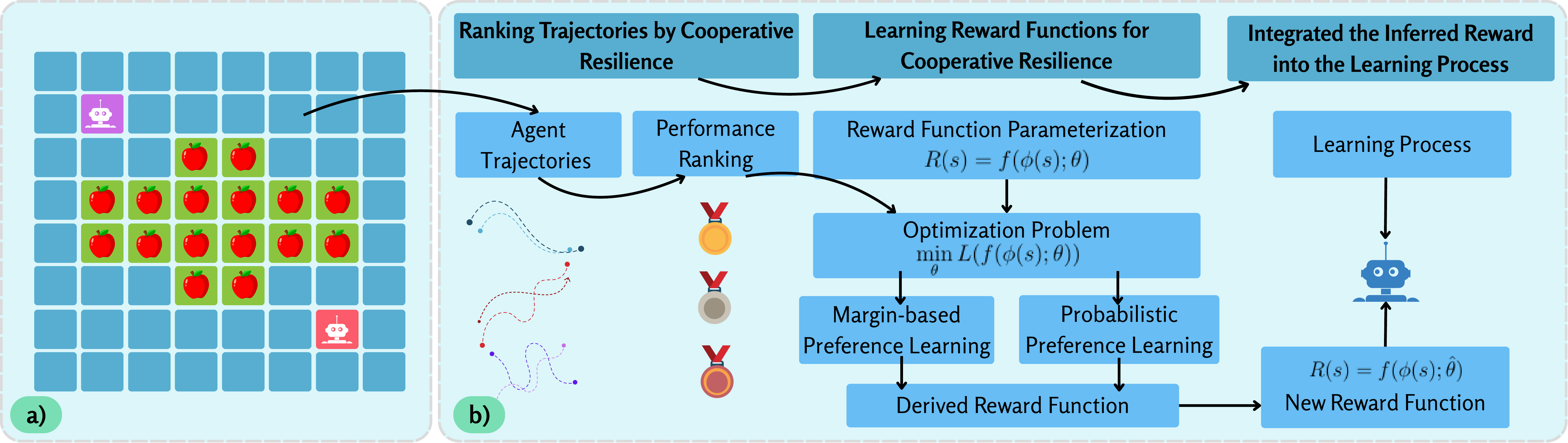}
    \caption{\textbf{(a)} Mixed-motive environment used throughout this study. Two agents interact in an \(8 \times 8\) grid with a central apple tree containing 16 apples. \textbf{(b)} Overview of our proposed reward learning pipeline.    
This figure illustrates the full loop from data collection to policy learning.}
    \label{fig:diagrama-enviroment}
    \vspace{-0cm}
\end{figure*}

Fig.~\ref{fig:diagrama-enviroment} summarizes the proposed learning pipeline. The process begins with a system of interacting agents operating in a mixed-motive environment, as illustrated in panel~(a). In this example setup, agents harvest resources from a shared apple tree in a grid-world, balancing individual consumption with long-term sustainability. Panel~(b) presents the reward learning pipeline. First, agent trajectories are collected and evaluated using a cooperative resilience metric. This evaluation induces a ranking over trajectories based on their resilience scores. Next, this ranking is used as input to a preference-based IRL module that learns a reward function aligned with resilient behaviors. The learned reward is then integrated into the agents' policy learning process, guiding behavior in future interactions with the environment. 

\section{Modeling Cooperative Resilience}
\label{sec:resilience}
\subsection{Definition of Cooperative Resilience}

Cooperative resilience is defined as the ability of a multi-agent system to sustain collective performance under perturbations through the coordinated behavior of its agents \cite{chacon2024cooperative, shraga2025collaboration}. It reflects how effectively a group anticipates, resists, adapts to, and recovers from disruptions that threaten its collective well-being. From a dynamical perspective, cooperative resilience captures two fundamental aspects of system response: the ability to absorb performance degradation during disruption and the ability to recover and restore functionality over time. As such, it provides a system-level characterization of how collective behavior evolves in the presence of uncertainty and adverse conditions.

\subsection{Trajectory-Level Evaluation}

To evaluate cooperative resilience, system behavior is evaluated at the trajectory level. A trajectory $\tau = (s_0, a_0, s_1, a_1, \cdots, s_T)$ represents the sequence of states and joint actions generated by agents interacting with the environment over a time horizon $T$, where $s_t \in S$ and $a_t \in A$ denote the state and joint action at the time step $t$. Let $\mathcal{D}$ denote the set of trajectories obtained under a given policy.

Each trajectory is assessed through a set of indicators that capture different dimensions of collective well-being. In this work, four indicators are considered: \textbf{Cumulative consumption}, computed \emph{per agent} (yielding one curve per agent); \textbf{Resource availability}, measured as the total number of apples present in the environment; \textbf{Gini index} of the consumption distribution, capturing inequality; and a \textbf{Hunger index}, estimating the delay between successive accesses to resources. The indicators we employ should be viewed as an instantiation of a more general template: the framework remains applicable as long as practitioners define the dimensions of collective well-being that matter for their particular domain.

For each indicator $I_k(t)$, performance is evaluated under two conditions: a baseline scenario without disruption and a disrupted scenario. This comparison allows for the characterization of how the system evolves before, during, and after perturbations. In particular, three key phases are identified: the onset of disruption at time $t_d$, the point of maximum degradation at $t_f$, and the recovery phase ending at $t_r$.

This trajectory-level representation enables the analysis of how collective behavior responds to disruptions across multiple dimensions, providing the basis for quantifying cooperative resilience in the following subsection.

\subsection{Aggregation into a System-Level Metric}

Based on the trajectory-level evaluation, cooperative resilience is quantified by comparing system performance under disrupted and baseline conditions. For each indicator $I_k(t)$, two temporal profiles are defined: a failure profile, which captures performance degradation after the failure, and a recovery profile, which captures the system's ability to restore functionality.

Formally, the failure and recovery profiles are defined as:
\begin{equation*}
\text{FP}_k = \int_{t_d}^{t_f} 
\dfrac{I_k^{\text{disrupted}}(t)}{I_k^{\text{baseline}}(t)} \, dt,
\end{equation*}

\begin{equation*}
    \text{RP}_k = \int_{t_f}^{t_r} \dfrac{I_k^{\text{disrupted}}(t)}{I_k^{\text{baseline}}(t)} \, dt.
\end{equation*}

Let $\Delta t_f = t_f - t_d$ and $\Delta t_r = t_r - t_f$ denote the durations of the degradation and recovery phases. The resilience score for indicator $k$ is then defined as:
\begin{equation*}
\rho_k = \dfrac{t_d + \text{FP}_k \cdot \Delta t_f + \text{RP}_k \cdot \Delta t_r}{t_d + \Delta t_f + \Delta t_r}.
\end{equation*}

This formulation captures the system's ability to absorb degradation and its ability to recover performance over time. 

To obtain a system-level measure, the indicator-specific scores are aggregated using the harmonic mean:
\begin{equation*}
\rho(\tau) = \left( \frac{1}{K} \sum_{k=1}^{K} \frac{1}{\rho_k} \right)^{-1}.
\end{equation*}

This aggregation penalizes low-performing dimensions and ensures that resilience reflects balanced performance across indicators. The resulting score $\rho(\tau)$ provides a single interpretable measure of cooperative resilience at the trajectory level and enables the comparison and ranking of trajectories based on their resilience.

\section{Learning Incentive Structures from Behavior}
\label{sec:learning}

\subsection{From Trajectory Rankings to Incentives}

Given trajectory-level resilience scores, a preference ordering over behaviors is obtained. Specifically, a trajectory $\tau_i$ is preferred over $\tau_j$ if $\rho(\tau_i) > \rho(\tau_j)$, where $\rho(\tau)$ denotes the cooperative resilience score. These preferences provide a behavioral signal that reflects desirable collective outcomes. Rather than manually specifying reward functions, we leverage this ranking to infer incentive structures that align agent behavior with cooperative resilience.

Formally, the objective is to learn a reward function $\hat{R}: S \rightarrow \mathbb{R}$ such that trajectories with higher resilience accumulate higher total reward. This establishes a direct link between the observed system-level performance and the incentives that guide individual decision-making.

\subsection{Incentive Function Parameterization}

To model incentive structures, the reward function is parameterized using a representation of the state characteristics, denoted as $\phi(s): S \rightarrow \mathbb{R}^n$. The learned reward is expressed as $R(s;\theta)$, where $\theta$ are the parameters to optimize.

Three parameterizations are considered:

\textbf{Handcrafted linear functions:} $R(s) = \phi(s)^\top w + b$, where $\phi(s)$ is designed manually to capture relevant system properties such as resource availability or fairness. This approach is interpretable, but depends on the quality of the feature.

\textbf{Linear functions based on the state:} $\phi(s) = s$, directly using the raw state variables. This avoids feature engineering, but may fail to capture complex relationships.

\textbf{Nonlinear models:} Neural networks $R(s;\theta)$ allow modeling nonlinear dependencies between states and resilience outcomes. These models are expressive, but require more data and introduce optimization challenges.

The choice of parameterization defines the hypothesis space over which incentive structures are learned.

\subsection{Preference-Based Learning Mechanisms}
\label{suc:preference}
Given a set of ranked trajectories, the learning problem is formulated using preference-based reinforcement learning. Two approaches to learning a reward function from these preferences: i) Margin-based Preference Learning (MPL) and ii) Probabilistic Preference Learning (PPL).

\subsubsection{Margin-Based Preference Learning}

This approach enforces that more resilient trajectories receive higher cumulative reward. Given a pair $(\tau_i, \tau_j)$ such that $\tau_i \succ \tau_j$, the objective is:
\[
\sum_{s \in \tau_i} R(s;\theta) > \sum_{s \in \tau_j} R(s;\theta).
\]

A margin $\delta_{ij}$ is introduced, fixed or proportional to the resilience gap $\delta_{ij} = |\rho(\tau_i) - \rho(\tau_j)|$. The optimization problem is:
{
\footnotesize
\begin{equation*}
\label{eq:hinge_loss}
\min_{\theta}  \quad \sum_{(\tau_i \succ \tau_j)} \max\left(0, \delta_{ij} - \left( \sum_{s \in \tau_i} R(s;\theta) - \sum_{s \in \tau_j} R(s;\theta) \right) \right).
\end{equation*}
}

\subsubsection{Probabilistic Preference Learning}

This approach models preferences probabilistically. The probability that $\tau_i$ is preferred over $\tau_j$ is defined using their cumulative rewards. The objective minimizes the negative log-likelihood:

{
\footnotesize 
\begin{equation*}
\label{eq:bt_loss}
\min_{\theta} \quad - \sum_{(\tau_i \succ \tau_j)} \log \left( \frac{\exp\left(\sum_{s \in \tau_i} R(s;\theta)\right)}{\exp\left(\sum_{s \in \tau_i} R(s;\theta)\right) + \exp\left(\sum_{s \in \tau_j} R(s;\theta)\right)} \right).
\end{equation*}
}

This formulation is convex when $R(s,\theta)$ is a linear function of state features, enabling efficient optimization using standard convex solvers. Compared to MPL, it provides smooth gradients, which can improve convergence stability and robustness to noisy or uncertain preferences. However, it may incur higher computational cost due to the exponential operations evaluated over trajectory pairs.

In both approaches, different data sampling strategies were evaluated to construct preference pairs (see Supplementary File Section A.4). These include random sampling, ranking-based sampling, and a mixed strategy combining both. In the margin-based formulation, this leads to six variants, obtained by combining two margin definitions ($\delta_{ij}=1$ and $\delta_{ij}=|\rho(\tau_i)-\rho(\tau_j)|$) with the three sampling strategies. Full implementation details are provided in the Supplementary File.

\section{Experimental Setting: Social Dilemma Environment}
\label{sec:experimental}

\subsection{Environment Description}
\label{sec:env-description}

We evaluate our approach in a simplified version of a \textit{social dilemma} inspired by the ``Commons Harvest'' scenario from the Melting Pot suite \cite{agapiou2022melting, perolat2017multi}. The original environment is defined as partially observable. In this work, we deliberately adopt a \emph{fully observable} variant. This is consistent with many fully observable Markov games used in cooperative and mixed-motive MARL. The environment consists of a discrete $8 \times 8$ grid where 2 agents interact and harvest resources from a shared tree containing 16 apples, located in the central region of the grid (see Fig.~\ref{fig:diagrama-enviroment}). Apples grow probabilistically, with regrowth chances increasing as more apples are preserved—encouraging sustainable behavior. This creates interdependence: while agents benefit from consumption, overharvesting reduces future availability and harms collective outcomes. This setting naturally captures a \textbf{social dilemma scenario}: agents must balance individual consumption with the long-term collective benefit of preserving the resource pool.

\subsection{Trajectory Collection and Resilience-Based Ranking}
\label{subsec:ranking}

To initialize the reward inference process, we collect 500 trajectories generated by agents following a random policy, each lasting 1000 steps. At timestep 500, a disruption removes apples from the central tree with fixed probability, ensuring that at least one remains so the episode can continue. These trajectories are then ranked according to their \textbf{cooperative resilience} score. The resulting ranking serves as input to preference-based algorithms to infer a reward function that promotes cooperative resilience. Full details of the experimental setup and configurations are provided in the Supplementary File Section A).

Note that the cooperative-resilience score is used  to construct the trajectory rankings and, as discussed later, as one of our evaluation metrics. This does not introduce direct circularity, since agents never receive the resilience score during training and are instead guided only by the inferred reward function, which is a function of the state rather than of the trajectory-level metric.

\subsection{Incentive Structures Evaluated}
\label{suc:incentive}
Building on the trajectory ranking, we evaluate multiple reward structures to determine which best support collective well-being. We consider two main approaches: i) resilience-based, where a unique reward function $\hat{R}$ is inferred and shared by all agents, and ii) hybrid, where $\hat{R}$ is combined with a consumption-based individual reward. In the hybrid setting, each agent receives a reward composed of the shared resilience term and its own consumption signal.

Both approaches are tested with three parameterizations of $\hat{R}$: handcrafted, state-based linear, and neural network, and the two optimization methods: MPL and PPL (see Subsection~\ref{suc:preference}). MPL yields six variants, from two margin schemes and three sampling strategies, while PPL yields three, resulting in 27 total configurations (see Supplementary File Section A.6).

\begin{figure}[b]
    \centering
    \begin{subfigure}[b]{0.45\textwidth}
        \centering
        \includegraphics[width=\textwidth]{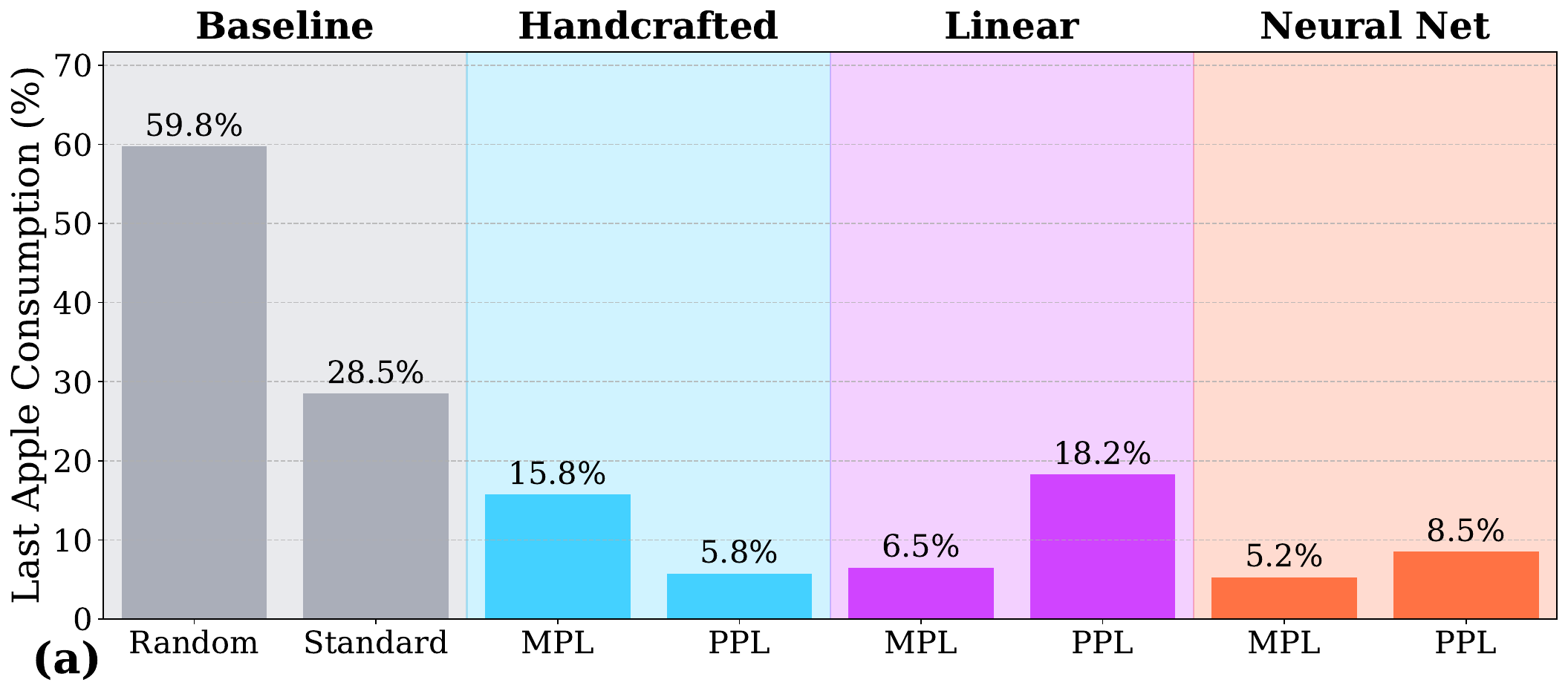}
    \end{subfigure}
    \hfill
    \begin{subfigure}[b]{0.45\textwidth}
        \centering
        \includegraphics[width=\textwidth]{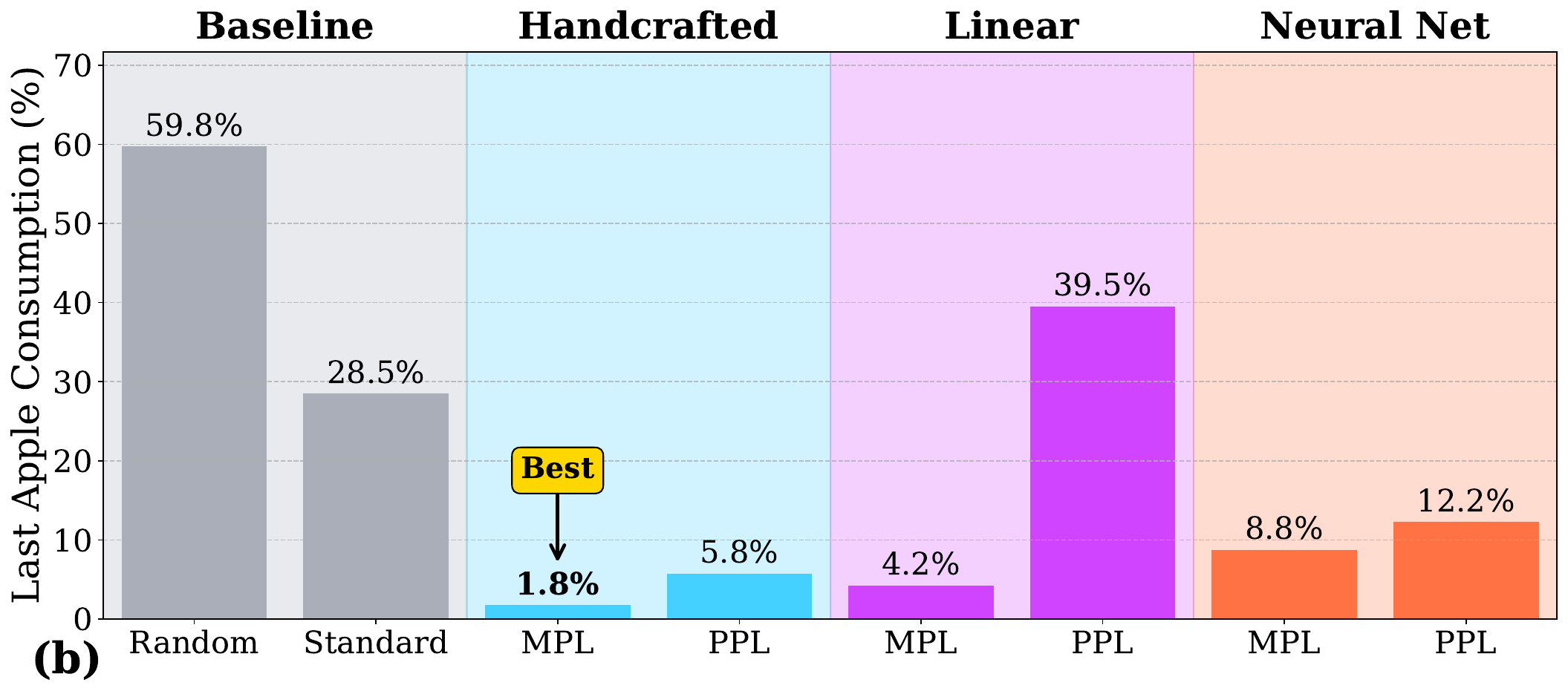}
    \end{subfigure}
    
    \caption{
    Percentage of episodes (out of 500) in which agents consumed the last remaining apple for the best configurations under each reward parameterization and optimization method. (a) Agents trained exclusively with resilience-aligned rewards. (b) Agents trained with \textit{hybrid incentive structure}.
    }
    \label{fig:last_apple_barplot}
\end{figure}

The inferred rewards are incorporated into a Proximal Policy Optimization (PPO) framework. Policies are trained for 500 episodes of 1000 steps each. The evaluation follows the same disruption protocol used during the trajectory ranking, where a subset of apples is removed from the central tree at step 500. Fig.~\ref{fig:last_apple_barplot} reports the metric associated with the consumption of the last remaining resource, comparing the best configurations of each reward parameterization and both optimization approaches. 

In selecting these \textit{best} configurations, we did not rely on resilience alone. Several PPL variants achieve high resilience scores (e.g., PPL-R and PPL-K), but do so by inducing overly conservative policies with low average rewards, preserving resources at the cost of individual task performance. To avoid such undesirable behaviors, we adopted a \textbf{multi-criteria selection rule}: high resilience, high cumulative reward, low last-apple consumption, and low variance across episodes. Under this joint evaluation, the MPL–M1 Hybrid with handcrafted features consistently dominated the alternatives. We adopt this reward configuration, hereafter referred to as our \textit{hybrid incentive structure}, as the reference for comparison against baseline methods under the new disruption protocol introduced in Section~\ref{sec:results}. Supplementary File Section A.5.3 provides additional metrics and plots for all configurations. 

It is important to note that our chosen \textit{hybrid incentive structure} relies on a handcrafted reward parameterization. Its strong performance is partly explained by the fact that its features encode meaningful prior structure about the domain, making it the least general and the most dependent on expert knowledge. At the same time, our preference-based IRL procedure must still learn appropriate weights for these features to obtain resilient, low-selfishness behavior; without the resilience-based rankings, the same features do not automatically yield suitable policies. By contrast, the linear and neural models operate directly over the full joint state and are therefore more data-hungry; with only 500 ranked trajectories in a high-dimensional, spatially structured environment, they are likely underpowered rather than fundamentally flawed.

\section{Results: Emergent Collective Behavior under Disruptions}
\label{sec:results}

We have implemented an expanded disruption protocol for evaluation, applied to agents trained with the \textit{hybrid incentive structure}. Evaluating algorithms under the same disruption protocol used during training may risk overfitting to known conditions, thus limiting the generalization claims of our method. To address this, the new protocol introduces three temporally distributed and qualitatively distinct disruptions, each lasting 5000 steps: i) resource removal at step 1250, ii) a temporary reduction in apple regrowth rate starting at step 2500, and iii) an agent failure simulation, where one agent loses control and moves randomly from steps 3750 to 3900.

In these evaluations, we consider three baselines: a random policy, PPO with a standard reward scheme (+1 for consuming an apple, 0 otherwise), and QMIX. For QMIX, the individual reward function is also defined using the +1/0 scheme, but in practice this leads agents to converge toward regions without apples. To mitigate this, we increased the reward to +10 for consuming an apple. Using this modified reward, the trained QMIX agents were evaluated with $\epsilon = 0$ under the disruption protocol. These baselines are contrasted against our \textit{hybrid incentive structure}, implemented as PPO with the resilience-informed reward learned through our method.

\subsection{System-Level Performance}

\begin{figure}[ht]
    \centering
    \begin{subfigure}[b]{0.43\textwidth}
        \centering
        \includegraphics[width=\textwidth]{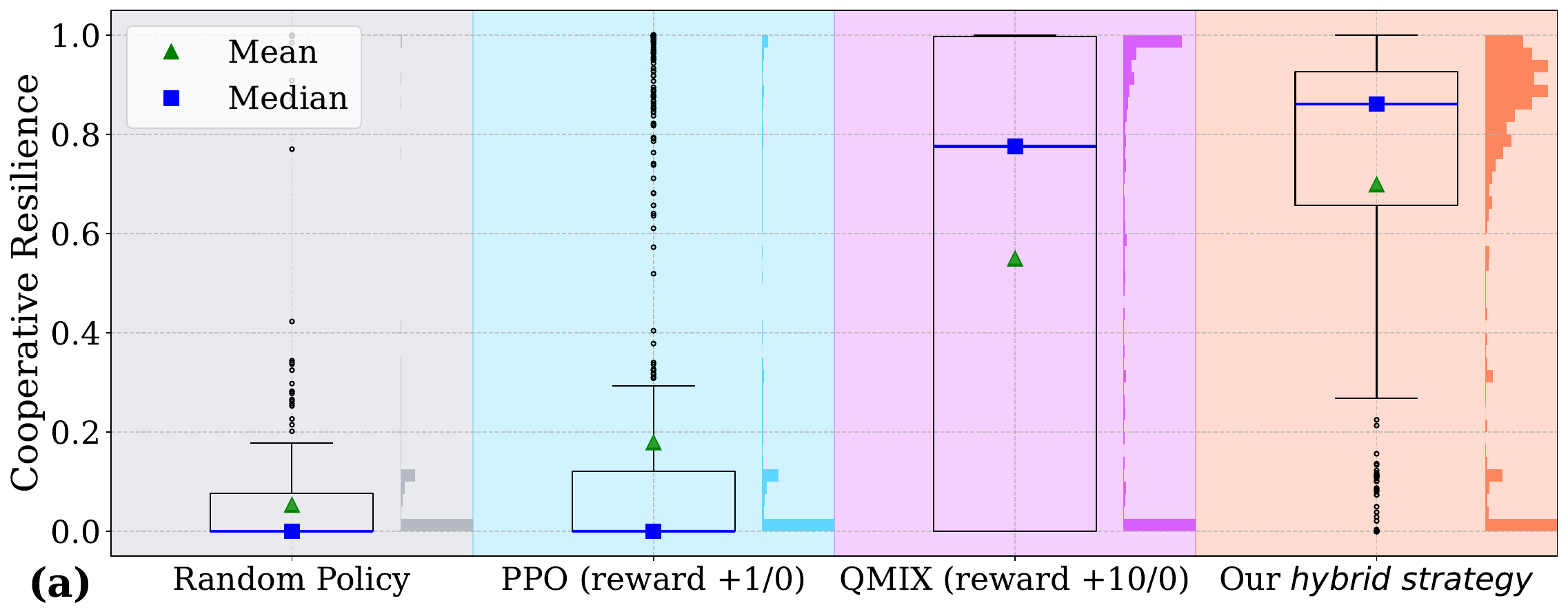}
       
    \end{subfigure}
    \hfill
    \begin{subfigure}[b]{0.43\textwidth}
        \centering
        \includegraphics[width=\textwidth]{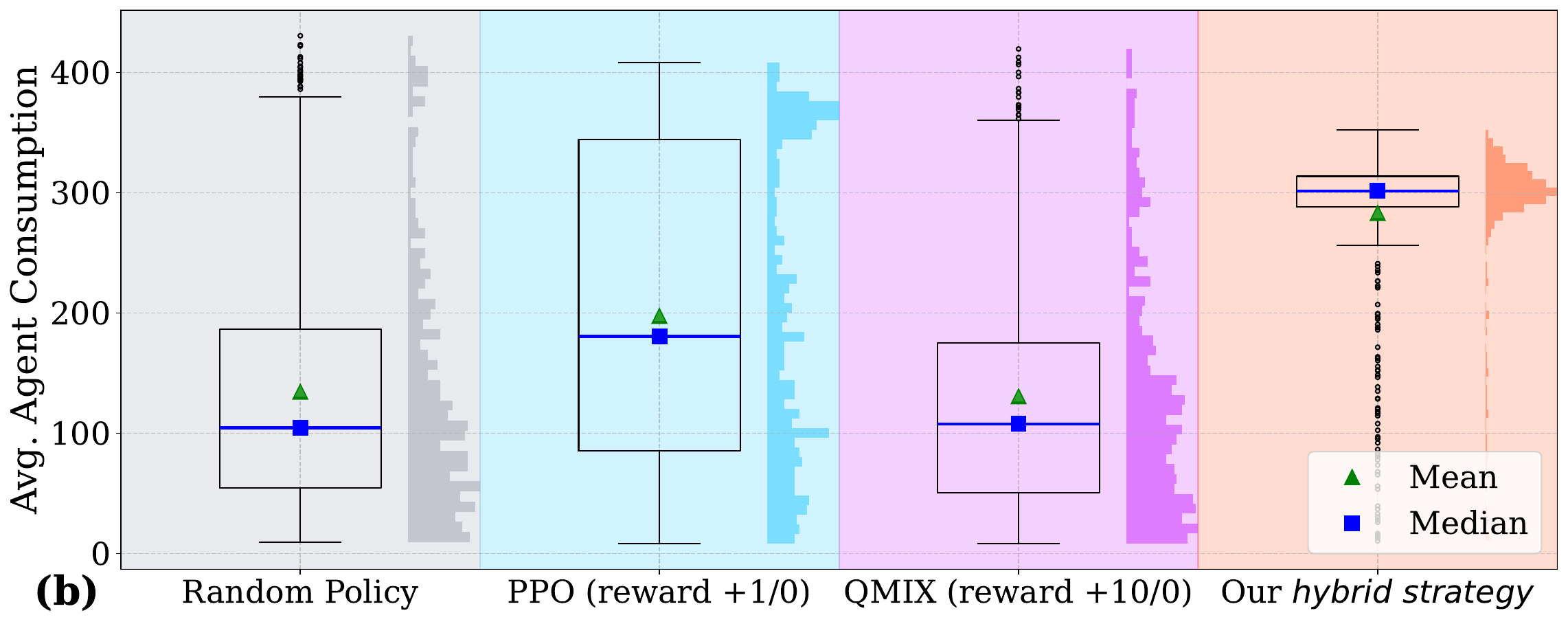}
     
    \end{subfigure}

    \begin{subfigure}[b]{0.43\textwidth}
        \centering
        \includegraphics[width=\textwidth]{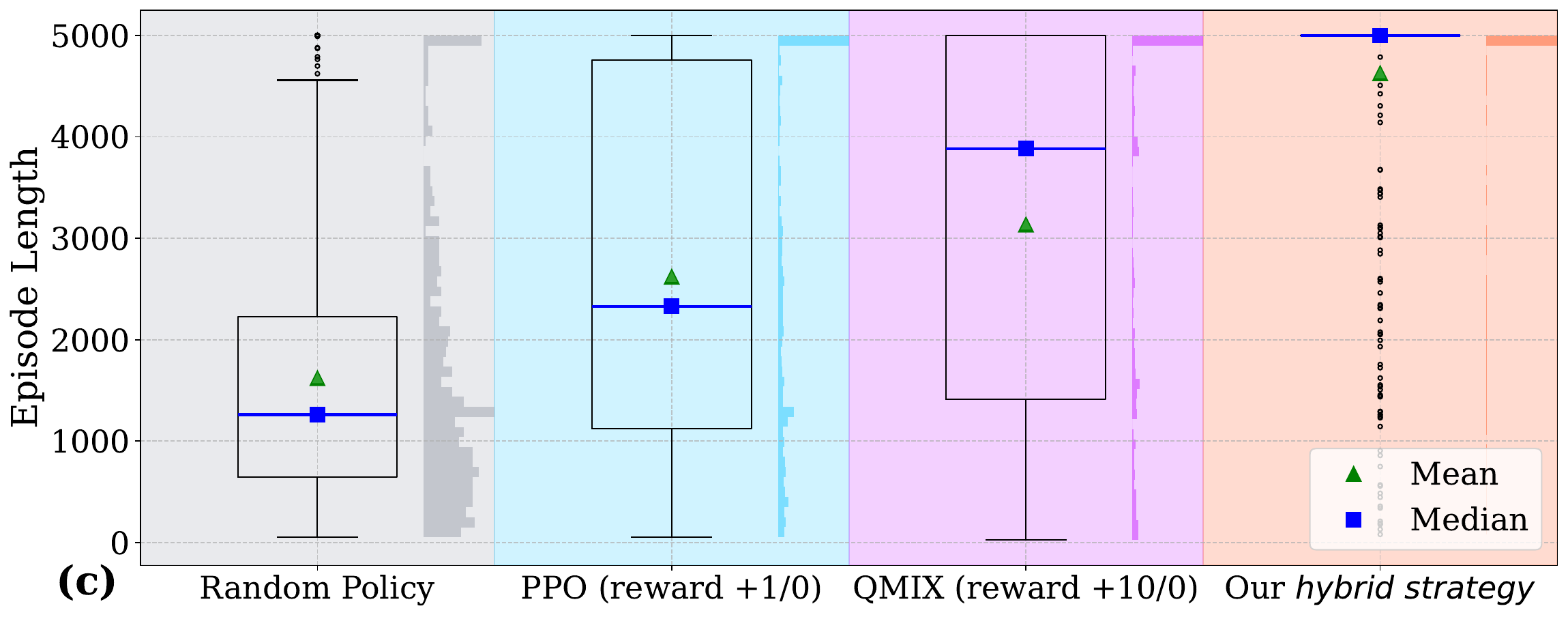}
       
    \end{subfigure}
    \hfill
    \begin{subfigure}[b]{0.43\textwidth}
        \centering
        \includegraphics[width=\textwidth]{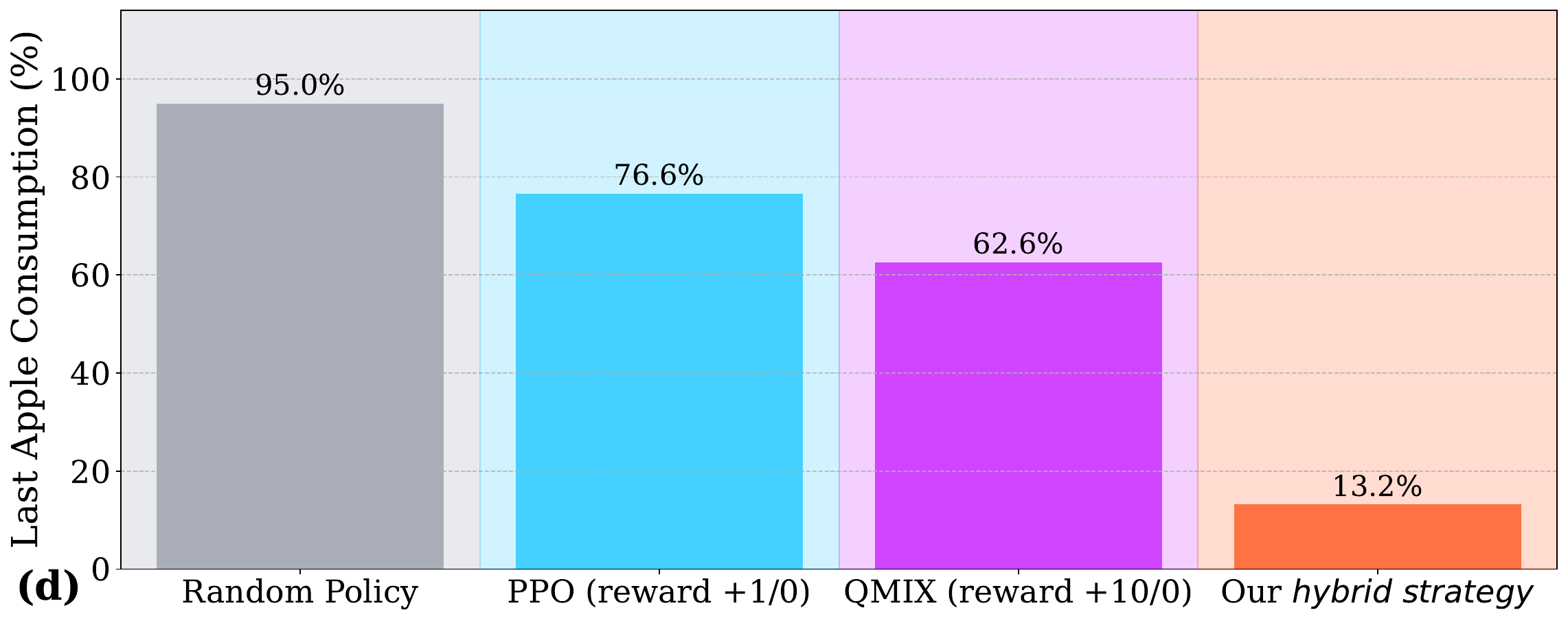}
    \end{subfigure}
    
    \caption{
    Performance comparison across 500 evaluation episodes under the expanded disruption protocol. (a) Cooperative resilience scores obtained by each method. (b) Average cumulative apple consumption per episode across both agents, reflecting task performance and resource utilization. (c) Episode duration, reflecting the ability of agents to sustain resource availability over time. (d) Frequency of last-apple consumption events, used as an indicator of social dilemma collapse caused by resource depletion. Boxplots report the distribution of results, while markers denote mean and median values.
    }
    \label{fig:metrics_evaluation}
\end{figure}

Fig.~\ref{fig:metrics_evaluation} summarizes system-level performance across 500 evaluation episodes. Panel (a) shows that cooperative resilience is consistently higher for the \textit{hybrid incentive structure}, with both mean and median values shifted toward the upper end of the scale relative to all baselines. Panel (b) indicates that the \textit{hybrid incentive structure} achieves the highest average cumulative consumption across agents, demonstrating that sustainability is attained without sacrificing productivity. Panel (c) further reveals that episode lengths are significantly extended. Under the \textit{hybrid incentive structure}, resources typically remain available until the simulation horizon (5000 steps), indicating more balanced and efficient resource utilization. These results suggest that the learned incentive structure supports sustained system performance over time.

To statistically validate these findings, we applied the Mann–Whitney U test for each metric, with $p$-values corrected using the Benjamini–Hochberg procedure (FDR, $\alpha=0.05$). Bonferroni-adjusted $p$-values are reported in Supplementary File Section A.8.1. Results confirm that, for cooperative resilience, \textit{hybrid incentive structure} significantly outperforms Random and PPO, while no significant difference is found relative to QMIX. In contrast, for both cumulative consumption and episode length, \textit{hybrid incentive structure} outperforms \textbf{all} baselines after correction.

\subsection{Collapse Analysis in Social Dilemmas}

A key failure mode in social dilemmas is the depletion of shared resources. Panel (d) in Fig.~\ref{fig:metrics_evaluation} captures this phenomenon through the last-apple consumption metric. The results show that the \textit{hybrid incentive structure} substantially mitigates collapse, with the last resource consumed in only 13.2\% of episodes, compared to significantly higher rates in baseline methods. This reduction indicates that agents trained with resilience-aligned incentives avoid overexploitation and maintain resource availability, preventing the system from entering irreversible failure states.

\subsection{Behavioral Patterns and Coordination}

\begin{figure*}[ht]
    \small
    \centering
    \includegraphics[width=\linewidth]{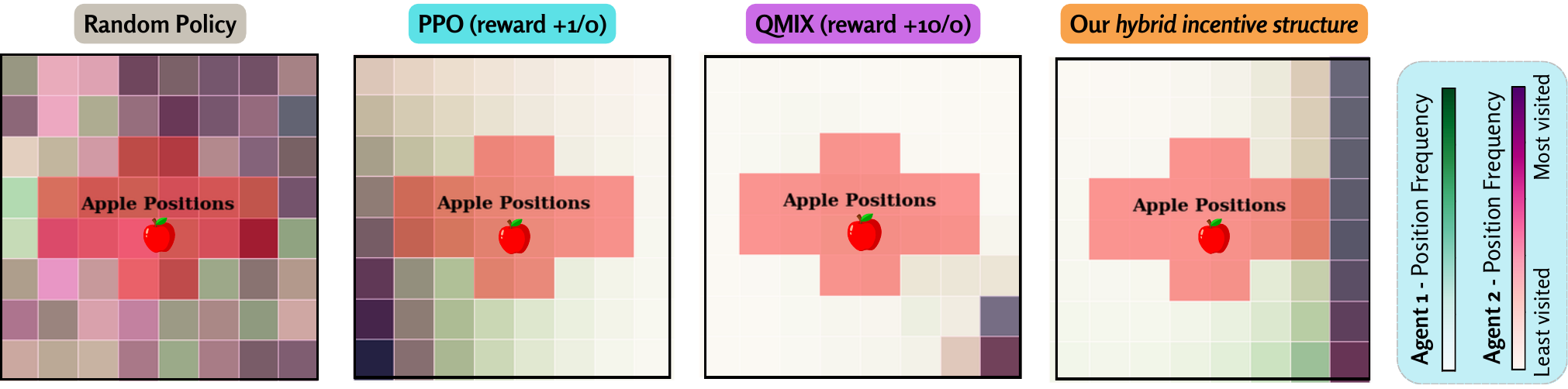}
    \caption{Position frequency maps for Agent~1 (green) and Agent~2 (purple) under four training configurations: (i) random policy, (ii) PPO with standard rewards, (iii) QMIX, and (iv) \textit{hybrid incentive structure}. Each heatmap depicts the spatial visitation density over 500 evaluation episodes, with apple locations marked in red. Agents were randomly initialized at the start of each episode and evaluated under the same protocol with three disruption events.}
    \label{fig:position_maps}
\end{figure*}

To further interpret agent behavior, we visualize position frequency maps over 500 evaluation episodes (Fig.~\ref{fig:position_maps}). Agent~1 is shown in shades of green and Agent~2 in purple, with apple positions marked in red. Under random policy, agents spread almost uniformly across the grid, with no clear coordination. With PPO rewards, both agents cluster in the bottom-left corner, strongly overlapping and competing for the same apples. QMIX produces an alternative but still suboptimal pattern: agents remain concentrated in the opposite corner without diversifying their movement across the grid. By contrast, the \textit{hybrid incentive structure} shows a complementary specialization: Agent~1 explores a wider area, while Agent~2 remains anchored along the right boundary, harvesting resources with little movement. This emergent division of roles could avoid redundant visits and illustrates how cooperative behavior can arise from differentiated strategies. For additional visualization, see Supplementary File Section C includes disaggregated position maps. These plots represent, with circles, the locations where each agent spent the most time, together with individual maps per agent to highlight their distinct spatial behaviors.

\begin{figure}[h]
    \centering
    \includegraphics[width=0.7\linewidth]{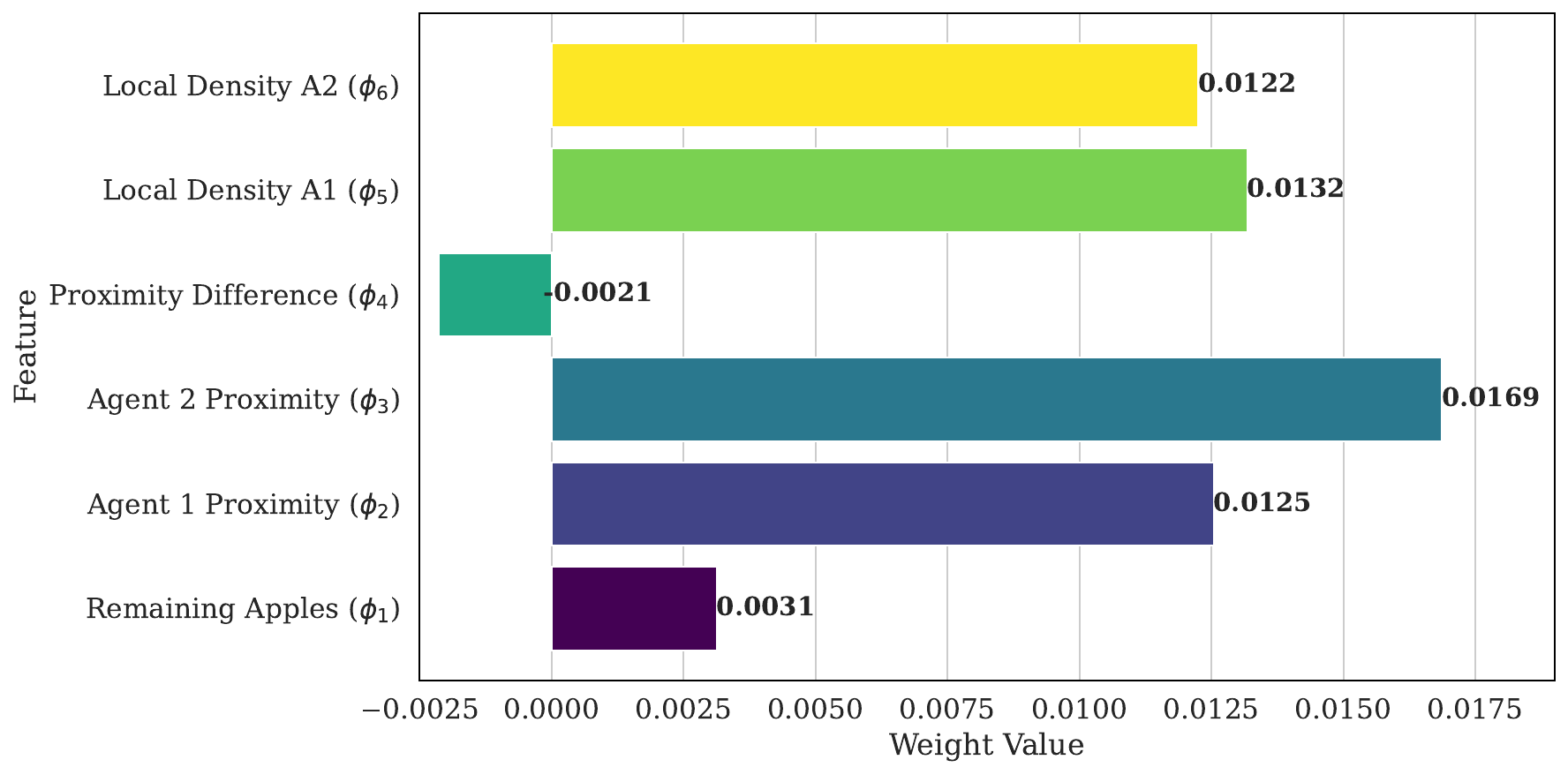}
    \caption{Learned weights associated with the handcrafted incentive structure in the best-performing configuration. The handcrafted reward function is parameterized using six interpretable features related to resource availability and spatial interaction patterns between agents. Although the weights should not be interpreted as absolute measures of feature importance, their relative magnitudes provide qualitative insight into the behavioral incentives induced by the learned reward function.}
    \label{fig:weights}
\end{figure}

To further analyze the behaviors induced by the handcrafted incentive structure, Fig.~\ref{fig:weights} presents the learned weights associated with the best-performing configuration. The handcrafted reward function is defined on six interpretable features related to collective resource usage and spatial interaction patterns: ($\phi_1$) the total number of remaining apples in the environment, ($\phi_2, \phi_3$) the distance of each agent to the nearest apple, the asymmetry between these distances ($\phi_4$) and ($\phi_5, \phi_6$) the local density of apples surrounding each agent.

The learned weights provide information on the incentives that shape collective behavior. The high positive weights associated with proximity and local density features suggest that agents are encouraged to remain in productive regions while avoiding direct competition over individual resources. In contrast, the negative weight associated with proximity asymmetry may indicate the emergence of complementary spatial roles between agents. This interpretation is consistent with the coordinated behaviors observed in Fig.~\ref{fig:position_maps}, where one agent explores broader regions of the environment while the other remains localized near resource-rich areas. Additional details on the handcrafted features and the interpretation of the learned weights are provided in Supplementary File Sections A.6 and B.

\subsection{Evaluating Scalability}

The initial evaluation of our approach was conducted in a simplified setting with few agents. Several extensions remain to be explored, including more complex and partially observable domains. To assess the scalability of our pipeline, we implemented a larger $16 \times 16$ grid-world with 4 agents and 3 apple trees (see Supplementary File Section A.8.2). In this environment, each tree disappears permanently once all surrounding apples are harvested, introducing a localized resource depletion mechanism and stronger interdependencies among agents. Moreover, resources can only regenerate up to a much lower threshold (16 apples in total, instead of the initial distribution), effectively limiting regrowth to the equivalent of a single tree. 

We applied our full pipeline in this extended environment, including the computation of resilience indicators and reward inference using \textit{hybrid incentive structure}. For reward inference, we relied on 400 ranked trajectories generated from random agent behavior (see Supplementary File Section A.7). The evaluation protocol consisted of 50 episodes, each lasting 2000 steps, with a disruption introduced in 300 timestep by removing apples from the environment. In this setting, we directly transferred the \textit{hybrid incentive structure} process identified in the smaller environment. Thus, the results here should be interpreted as a practical workflow in which configuration searches are performed in small environments and then the scalability of the larger ones is evaluated.

The algorithm was compared against a random policy and PPO with the traditional reward scheme. As summarized in Table~\ref{tab:comparison_16_16}, our \textit{hybrid incentive structure} achieves higher average cooperative resilience, longer episode durations, and a reduction in social dilemma failures. To further evaluate robustness, we applied the Mann–Whitney U test with Benjamini–Hochberg corrections when comparing \textit{hybrid incentive structure} against standard PPO. The results suggest that, although the improvement in average resilience is not statistically significant in this setting, resilience-based rewards during training lead to significantly longer episodes and higher cumulative rewards. These findings indicate enhanced system survivability and sustained agent performance, consistent with the intended goals of cooperative resilience. Additional results and statistical analyses for this environment are provided in Supplementary File Section A.8.2.

\begin{table*}[ht]
\centering
\footnotesize
\caption{Comparison of algorithms in the extended 16$\times$16 environment with four agents.}
\label{tab:comparison_16_16}

\begin{tabularx}{\textwidth}{lcccc}
\toprule
\textbf{Method} & 
\textbf{Cooperative Resilience} & 
\textbf{Apple Consumption} & 
\textbf{Episode Length} & 
\textbf{Last Apple} \\
\hline \midrule
\textit{Hybrid incentive structure} & $0.889 \pm 0.395$ & $22.93 \pm 4.77$ & $1923 \pm 263$ & 6 / 50 \\
PPO (reward +1/0) & $0.814 \pm 0.469$ & $16.74 \pm 4.50$ & $1450 \pm 625$ & 25 / 50 \\
Random policy & $0.274 \pm 0.293$ & $14.51 \pm 3.47$ & $760 \pm 361$ & 50 / 50 \\
\bottomrule
\end{tabularx}
\end{table*}

\section{Discussion of Results}
\label{sec:discussion}

\subsection{Incentives and Collective Behavior}

The results show that cooperative resilience can be improved through incentive structures learned from behavior. Agents trained with the \textit{hybrid incentive structure} consistently outperform baseline policies, achieving higher resilience and more structured interaction patterns, even when trained from random demonstrations. The proposed incentive structure balances individual and collective objectives. This balance allows agents to maintain task performance while promoting cooperative outcomes. In particular, the frequency of consumption of last-apples, used as a proxy for selfish behavior, drops to 13.2\%, while the total consumption of resources remains higher than in the baseline methods. This indicates that agents avoid collapse without sacrificing productivity.

Behavioral analysis further supports these findings. Spatial maps reveal complementary roles where one agent explores the environment, while the other remains localized in resource-rich areas. This division of behavior reduces competition and supports the sustainable use of resources. These patterns are consistent with the structure of the learned incentives, which promote coordination without requiring explicit communication.

Overall, these results highlight the role of incentive design as a mechanism for shaping collective behavior and enabling resilient cooperation in multi-agent systems.

\subsection{Sensitivity to the Resilience Metric}

The proposed framework relies on a resilience metric defined through multiple indicators of collective well-being. Although this design provides a flexible and interpretable evaluation, it also introduces sensitivity to the choice of indicators and their relative importance. In this work, the metric includes consumption, resource availability, inequality, and access delay. These indicators capture key aspects of social dilemmas, but alternative domains may require different definitions of collective well-being. As a result, the learned incentive structures are inherently tied to the selected metric. The framework allows practitioners to encode domain-specific notions of desirable behavior through the choice of indicators. However, it also implies that careful design and validation of the resilience metric are essential to ensure meaningful outcomes. 

\subsection{Scalability and Computational Aspects}

Scaling the proposed framework to larger systems introduces both computational and methodological challenges. Resilience evaluation requires computing trajectory-level indicators under baseline and disrupted conditions, which increases the cost of data collection and processing. The reward learning stage further adds complexity, as preference-based learning relies on pairwise trajectory comparisons, leading to quadratic growth in the number of training samples. This effect becomes more pronounced as the number of agents and the size of the environment increase.

Despite these challenges, the results in the extended $16 \times 16$ environment suggest that the framework can be used to configurations with more agents and more complex interaction dynamics. The learned incentive structures improve the duration of the episode and reduce collapse events, indicating improved system survivability.

However, extending the framework to partially observable environments introduces additional difficulties. In its current form, resilience evaluation assumes access to the full state of the system, effectively relying on an oracle metric. This constitutes a strong assumption that may not hold in realistic settings. One possible extension is to estimate the resilience based on the beliefs of the agents about the state. Although this approach could relax the observability requirement, it would likely increase computational cost and introduce additional sources of uncertainty.

\subsection{Limitations}

This study has some limitations. First, the empirical evaluation is restricted to a limited set of environments and baseline methods, including random policies, PPO, and QMIX. More recent cooperative MARL approaches were not included in the comparison. However, it is important to note that the proposed framework does not aim to introduce a new multi-agent training algorithm. Instead, it provides a mechanism for learning incentive structures that can, in principle, be integrated into a wide range of MARL methods that rely on reward specification.

Second, the framework assumes full observability and discrete action spaces. Extending the approach to partially observable environments or continuous domains remains an open challenge. In particular, resilience evaluation in its current form requires access to the full state of the system. Third, the learned incentive structures depend on the availability of sufficient trajectory data. In complex environments, collecting, evaluating, and ranking trajectories may become computationally expensive, especially when disruptions must be simulated or repeated under controlled conditions.

In addition, the resilience evaluation assumes that the disruption onset time is known. This effectively introduces an oracle assumption, which may not be realistic in operational settings where perturbations are not explicitly observable. Relaxing this assumption would require an additional stage for detecting anomalies or identifying disruption events from observed data. Such extensions would introduce uncertainty into the evaluation process and may further increase the computational cost.

Addressing these limitations is an important direction for future work, particularly in the context of large-scale, partially observable, and real-world multi-agent systems operating in social dilemma settings.

\section{Conclusion}
\label{sec:conclusions}

This work introduced a framework for learning incentive structures from behavior in mixed-motive multi-agent systems. Using a cooperative resilience metric to rank trajectories, the proposed approach infers reward functions that align individual decision-making with desirable system-level outcomes. The results suggest that the learned incentive structures improve cooperative resilience, extend the sustainability of the system, and reduce collapse events, while maintaining effective resource utilization. In addition, the induced incentives lead to structured and coordinated behaviors, illustrating how collective outcomes can emerge from appropriately aligned individual rewards. This work presents an approach for identifying incentive structures that may be integrated into existing MARL methods. This perspective highlights the role of reward design as a key component in shaping resilient collective behavior in social dilemmas.

Future work will focus on extending the framework to partially observable and large-scale environments, as well as relaxing assumptions such as full state observability and known disruption times. Another promising direction is the incorporation of alternative sources of behavioral data, including human-generated trajectories, to further improve the robustness and applicability of learned incentive structures in real-world systems.





\bibliographystyle{IEEEtran}
\bibliography{references}

\end{document}